\documentclass[floatfix,showpacs,amsmath,amssymb,letterpaper,groupaddresses,superscriptaddress]{article}
\usepackage{graphicx}
\usepackage{latexsym}
\usepackage{amsmath}
\usepackage{amssymb}
\usepackage{epsfig}
\usepackage{graphicx}
\usepackage{graphicx}
\usepackage{amssymb,amsmath,times}
\usepackage{hyperref}
\usepackage{color}

\def\be{\begin{equation}}
\def\ee{\end{equation}}
\def\ba{\begin{eqnarray}}
\def\ea{\end{eqnarray}}
\def\la{\langle}
\def\ra{\rangle}

\def\h{\hskip 1cm}

\def\lo{\longrightarrow}

\def\u{\uparrow}
\def\d{\downarrow}

\usepackage{epsfig}
\usepackage{graphicx}
\begin{document}

\begin{center}
{\Large \bf A comparison of parallel and anti-parallel two qubit mixed states}\\

\vspace{1cm} Azam Mani,\hspace{5mm}
Vahid Karimipour, \hspace{2mm} and \hspace{2mm} Laleh Memarzadeh, 
\vspace{5mm}

Department of Physics, Sharif University of Technology,\\
P.O. Box 11155-9161,\\
Tehran, Iran.\\

\end{center}
\vskip 3cm

\begin{abstract}
We investigate the correlation properties of separable two qubit states with maximally mixed marginals. 
These stats are divided to two sets with the same geometric quantum correlation. However a closer scrutiny of these states reveals a profound difference between their quantum correlations as measured by more probing measures.
Although these two sets of states are prepared by the same type of quantum operations acting on classically correlated states with equal classical correlations, the amount of final quantum correlation is different. We investigate this difference and trace it back to the hidden classical correlation which exists in their preparation process. We also compare these states with regard to their usefulness for entanglement distribution and their robustness against noise.   
\end{abstract}


\section{Introduction}

Entanglement and superposition are distinctive quantum mechanical features which can be used to surpass the limitations of classical information 
processing \cite{tele, dense}. The physical and technological impact of these effects are so large that entanglement is considered as a resource, in the same way as energy. Entanglement is classified \cite{entClassification1, entClassification2}, quantified \cite{quantifyEnt, Wootters}, manipulated \cite{entManipulation1,entManipulation2} and distributed \cite{entDistribution1, entDistribution2}.  Even various types of networks of entangled states are being investigated \cite{entDistribution1, entDistribution2, network2}.  Like any other resource, it is questioned whether this is the only resource which we can rely on, or there are other cheaper and less fragile resources which can be equally effective in at least certain subclasses of our quantum communication tasks. It is now known that there are indeed separable states which do have some type of quantum correlations \cite{discord,Groisman,Modi,Gessner, LQU}. In recent years, the same type of study as mentioned above, has begun to emerge for these kinds of states \cite{Knill, Datta,Madhok, DiscInterpretation, Wang, RSP}.
  For example, questions like: how much correlation exist in a separable state, how such a correlation can be produced, how robust it is against noise, and whether or not this correlation can be distilled are of conceptual and practical relevance.  In this paper we want to investigate some of these questions for an important class of states, namely two qubit mixed states with maximally mixed marginals or the so-called Bell diagonal states. While we do a rather general study of these states and their  properties, we would like to emphasize the interplay of two specific factors, namely the method of preparation and the amount of quantum correlations in these states. To present the problem definitely, consider a smaller subclass, namely Werner states \cite{Werner} which are defined as:
\be\label{WernerAll}
W(t)=\frac{1-t}{4} I + t|\Psi^-\ra\la \Psi^-|,
\ee
where $|\Psi^-\ra$ is the singlet state $|\Psi^-\ra=\frac{1}{\sqrt{2}}(|01\ra-|10\ra)$. The importance of these states stems from the fact any two qubit state can be converted to a Werner state by bi-local unitary operations \cite{WernerImportance}. It is well known that such a state is separable when the parameter $t$ is restricted to the range $[-\frac{1}{3},\frac{1}{3}]$. \\

For concreteness, consider the case where $t=\frac{1}{3}$ and $t=-\frac{1}{3}$, where we denote the states $W(\frac{-1}{3})$ and $W(\frac{1}{3})$ respectively by $W^{\u\u}$ and $W^{\u\d}$. These states are separable and can be decomposed respectively as follows:
\begin{equation}\label{W'}
W^{\u \u}=\frac{1}{6} \sum_{n=x,y,z}\left( |n^\u,n^\u\ra\la n^\u,n^\u|+|n^\d,n^\d\ra\la n^\d,n^\d| \right)
\end{equation}
and
\begin{equation}\label{W}
W^{\u\d}=\frac{1}{6} \sum_{n=x,y,z}\left( |n^\u,n^\d\ra\la n^\u,n^\d|+|n^\d,n^\u\ra\la n^\d,n^\u| \right).
\end{equation}  
Therefore for $t=\frac{-1}{3}$, a Werner state is a uniform mixture of parallel spins and for $t=\frac{1}{3}$, it is a uniform mixture of anti-parallel spins. At a first glance, it seems that there is not much difference between the above two states. However, as we will see, these two states have different amount of quantum correlations, and have different efficiency in performing certain quantum communication tasks.  Thinking of them as resources, we may ask the following questions:\\

\begin{itemize}
\item{
 Which of the above two states has higher value of quantum correlation and what is the origin of it?}\\
\item{
 Which of the above states is more useful for quantum communication tasks?}
\item{
 Which one is harder to prepare by local operations and classical communication?}
\item{
 Which one is more robust under local noise?}

\end{itemize}

We will try to present comprehensive answers to these questions. To this end we study them in the more general setting of one-parameter family of Werner states which we write as follows:

\begin{equation}\label{Wuu}
W^{\u\u}(t)=\frac{1+t}{4}I -t|\psi_-\ra\la \psi_-|, \h 0\leq t\leq \frac{1}{3},
\end{equation}
and
\begin{equation}\label{Wud}
W^{\u\d}(t)=\frac{1-t}{4}I + t|\psi_-\ra\la \psi_-|, \h 0\leq t\leq \frac{1}{3}.
\end{equation}
Analyzing the one parameter family of states in (\ref{Wuu}) and (\ref{Wud}) gives us a comprehensive answer to the questions above. Note that our analysis is not restricted to the case of Werner states, in fact we study all separable two qubit states with maximally mixed marginals with regard to the above four questions and in some discussions we specifically analyze the one parameter family of (\ref{Wuu}) and (\ref{Wud}) in order to illustrate our results.\\

Stated briefly, we show that the set of separable two qubit states with maximally mixed marginals can be divided to two non-equivalent classes which are joint to each other in a measure zero subset. The states of one class are mixtures of maximally mixed state with parallel spins along the $x$, $y$ and $z$-axes while the states of the other subset are mixtures of maximally mixed state with anti-parallel spins along the same axes, hence we use the notation $\rho^{\u\u}$ and $\rho^{\u\d}$ to represent the corresponding elements of these classes respectively (consider (\ref{W'}) and (\ref{W}) as special cases).\\

After a short discussion about the importance of the preparation of separable quantum correlated states, we will present a preparation method for all such states and discuss that our method is more efficient than the recently proposed one \cite{generation}. We also find that the states $\rho^{\u\u}$ have more quantum correlation content than $\rho^{\u\d}$ as measured by local quantum uncertainty \cite{LQU}. This is intriguing in view of the fact that both these states are prepared by acting on two equally classically correlated states by identical quantum channels. We relate this difference to a hidden classical correlation which is needed for setting up aligned coordinate systems by the two parties in order to enact the correlated quantum channel. It turns out that for producing parallel states $\rho^{\u\u}$ a more precise alignment is necessary compared with the case when one wants to produce anti-parallel states $\rho^{\u\d}$. This extra classical correlation is what goes into the total quantum correlation of the parallel states.
Of course one can argue that the state $\rho^{\u\d}$ can be prepared (at less cost) and then covered to  $\rho^{\u\u}$ by an optimal NOT operation \cite{optNOT1}.  But as we will show, a successful conversion via the optimal NOT (with good fidelity) also amounts to having more initial classical correlations for preparation of $\rho^{\u\u}$.
It should be noted that while our preparation method requires a set of agreed up on coordinate axes by the two parties, it does not concern the aligning process of the axes. In fact there are different methods for setting up a reference frame (see the review article \cite{RF-review} and the references inscribed), for example two distant parties may use pure parallel or anti-parallel spin states to set up an agreed up on coordinate system \cite{Gisin}. Nevertheless we will discuss that our problem and the problem of reference frames concern different issues in the field of quantum information. Precisely, some literature of reference frames are about efficient use of quantum resources for transmission of a reference frame \cite{RF-review, Gisin, RF1} and some are about quantum communication without having a precise reference frame \cite{RF2}, while in this paper we do not use the states $\rho^{\u\u}$ and $\rho^{\u\d}$ for setting up a reference frame or for quantum communication between two distant parties.\\

%

\begin{figure}[tp]
\begin{center}
\vskip -3 mm
\includegraphics[width=12cm, height=1.5cm]{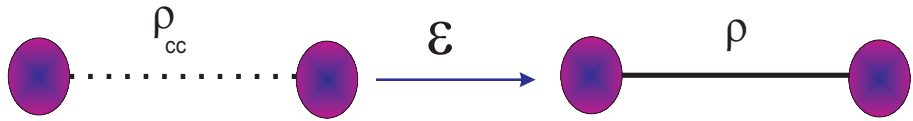}
\caption{(Color online) 
One can prepare separable states with quantum correlations by the action of bi-correlated unitary maps. For those states whose marginals are maximally mixed, only three types of unitary operators are necessary to produce all such states, as described in the discussion leading to  eq. (\ref{E}).}
\end{center}
\label{BasicChannel}
\end{figure}


The structure of this paper is as follows: in section (\ref{preliminary}) we review some preliminary facts about two-qubit separable states. In section (\ref{correlation}) we calculate and compare the quantum correlations of the desired states. 
We then present a preparation method for separable states in section (\ref{preparation}).
Equipped with these tools, in sections (\ref {origin}) we discuss about the origin of the correlation difference of the states $\rho^{\u\u}$ and $\rho^{\u\d}$. In section (\ref{optNotSection}) we present an alternative preparation method which is based on the action of the optimal NOT operator. Afterward in section (\ref{effect}) we compare the effectiveness of the states of the two classes in a quantum information task, more precisely we present an entanglement distribution protocol with Werner states. We also study the effect of depolarizing noise on these states in this section. Finally we end the paper with a discussion in section (\ref{discussion}) . \\

\section{Preliminary facts about two-qubit mixed states}\label{preliminary}
Since we want to analyze the set of two qubit states with maximally mixed marginals with regard to the questions we asked in the introduction,  in this section we berifely present some preliminary facts about such states. These states can be written as:
\be\label{MaxMix}
\rho=\frac{1}{4}(I+t_{ij}\sigma_i\otimes \sigma_j),
\ee
where $\sigma_i$ ($i=1,2,3$) are Pauli matrices and  summation over repeated indices is understood. By local unitary actions, the matrix $t_{ij}$ can be diagonalized:
\be\label{diagT}
\rho=\frac{1}{4}(I+\sum_{i=1}^3t_{i}\sigma_i\otimes \sigma_i).
\ee
Let us denote the space of such states by $\Lambda$. To ensure positivity, the parameters $t_i$ should be confined within a regular tetrahedron, whose vertices are given by the vectors
$$e_0:=(-1\ ,-1\ ,-1),\h e_1:=(-1\ ,1\ ,1\ ),\h e_2:=(1\ ,-1\ ,1),\h e_3=(1\ ,1\ ,-1).$$
Not all the states in this tetrahedron are separable. To be a separable state, the parameters $t_i$ should be restricted to a regular octahedron inscribed in the above tetrahedron. The vertices of this octahedron are given by
$$v_1^{\pm}:=(\pm 1\ ,0\ ,0),\h v_2^{\pm}:=(0\ ,\pm 1\ ,0\ ),\h v_3^{\pm}:=(0\ ,0\ ,\pm 1).$$

The local unitary action of part A (Alice) by a Pauli matrix $\sigma_1\otimes I$ changes the signs of $t_2$ and $t_3$ while leaving the sign of $t_1$ intact. A similar thing happens with other local Pauli operators. Therefore depending on the sign of $t_1t_2t_3$, $\Lambda$ is divided into two inequivalent classes of states denoted by $\Lambda^{\u\u}$ and $\Lambda^{\u\d}$. The representative elements of these classes are respectively as follows:

\be\label{rhouu}
\rho^{\u\u}= \frac{1}{4}(I+t_1\sigma_1\otimes \sigma_1+t_2\sigma_2\otimes \sigma_2+t_3\sigma_3\otimes \sigma_3), \h 0\leq t_1, t_2, t_3\leq 1,
\ee
and
\be\label{rhoud}
\rho^{\u\d}= \frac{1}{4}(I-t_1\sigma_1\otimes \sigma_1-t_2\sigma_2\otimes \sigma_2-t_3\sigma_3\otimes \sigma_3), \h 0\leq t_1, t_2, t_3\leq 1,
\ee
both subject to the condition $0\leq t_1+t_2+t_3\leq1$ (needed for positivity of the matrix). As we will show, the states (\ref{rhouu}) and (\ref{rhoud}) are constructed from a mixture of maximally mixed state and parallel ($\u\u$) or anti-parallel ($\u\d$) spins respectively, and this is the reason for the notation that we have used. Note that these two sets are joined to each other along a subset of measure zero, where $t_1t_2t_3=0$. Since any state of the form (\ref{MaxMix}) can be converted to (\ref{rhouu}) or (\ref{rhoud}) by local unitary actions, in order to study the correlation properties of general two-qubit states of the form (\ref{MaxMix}), we need only study the properties of these special classes.\\

First let us decompose these two states to a convex combination of pure states, this decomposition shows their difference in a transparent way and turns out to be important in our subsequent discussion. To do this, we define the pure states:

\be
P_i^{\pm}=\frac{1}{2}(I\pm \sigma_i),
\ee
and note that $\sigma_i\otimes \sigma_i$ can be written in two different ways in terms of product of these pure states, namely:
\be\label{sigma-uu}
\sigma_i\otimes \sigma_i=2(P_i^+\otimes P_i^++P_i^-\otimes P_i^-)-I,
\ee
or
\be\label{sigma-ud}
\sigma_i\otimes \sigma_i=I-2(P_i^+\otimes P_i^-+P_i^-\otimes P_i^+).
\ee
In order to write the states (\ref{rhouu}) and (\ref{rhoud}) as a convex combination of product states, we use one of the above formulas as appropriate. One finds that 

\be\label{uuDecom}
\rho^{\u\u}
=\frac{1}{4}\left((1-t_1-t_2-t_3)I+\sum_{i=1}^3 2t_i(P_i^+\otimes P_i^++P_i^-\otimes P_i^-)\right)
\ee
and
\be\label{udDecom}
\rho^{\u\d}
=\frac{1}{4}\left((1-t_1-t_2-t_3)I+\sum_{i=1}^3 2t_i(P_i^+\otimes P_i^-+P_i^-\otimes P_i^+)\right).
\ee
\\

Therefore $\rho^{\u\u}$ is a mixture of maximally mixed state with a convex combination of states of parallel spins along the three axes $x$, $y$ and $z$, while $\rho^{\u\d}$ is a mixture of maximally mixed states and a combination of anti-parallel spin states along the same axes.
When one of the parameters $t_i$ say $t_3=0$, the two states are locally convertible to each other, i. e. $(I\otimes \sigma_3)\rho^{\u\u}(I\otimes \sigma_3)=\rho^{\u\d}$. This is in fact due to the existence of a universal NOT operator for equatorial states, which is nothing but the $\sigma_3$ operator. This operator can easily reverse the orientation of any spin state in the equatorial plane: i.e.  $\sigma_3: |\phi\ra=\frac{1}{\sqrt{2}}(|0\ra+e^{i\phi}|1\ra)\lo \frac{1}{\sqrt{2}}(|0\ra-e^{i\phi}|1\ra)=|\phi^\perp\ra.$
The same is also true if any other parameters $t_1$ or $t_2$ are zero. However, when all the parameters are different from zero, the two states $\rho^{\u\u}$ and $\rho^{\u\d}$ are not exactly convertible to each other, due to the non-existence of a universal NOT operator. \\

{\bf Remark:} Although a universal NOT operation does not exist \cite{optNOT1}, one can come close to it with arbitrary fidelity. In fact in \cite{optNOT1} it is shown that such an optimal NOT operator can be constructed by first estimating a state from $N$ copies of a given state $\sigma$ (with fidelity $F=\frac{N+1}{N+2}$) and then preparing the complement  state $\sigma^\perp$. We will further discuss this in section (\ref{optNotSection}).\\

We will see that the two states ${\rho}^{\u\u}$ and ${\rho}^{\u\d}$, although very similar to each other, do not have the same performance in quantum information processing tasks. In fact this is due to the difference in their quantum correlation content. The origin of this difference is subtle and we will argue why this is so, after we have shown how these two states can be prepared from a classically correlated state. 
Before doing this we compare these states with regard to their quantum correlation content.

\section{Comparison of quantum correlations of the states}\label{correlation}

For a bipartite system, quantum correlation can be defined as the difference between the total and classical correlations \cite{discord}. Total correlation is equal to the mutual information of the bipartite system, and the classical correlation is defined to be the maximum amount of information that can be attained from the whole system by performing local measurements on one of the two subsystems \cite{discord}.
Naturally calculation of quantum correlation requires a formidable optimization which can be carried out only for a restricted class of states \cite{qubitdiscord}. As substitutes, other computable measures have been proposed in the literature. Some are based on geometric approaches and are defined as  the distance between a given state and the closest classically correlated state \cite{GeometricDiscord}, others are based on fidelity with such a state \cite{Bruss}, or on non-commutativity of reduced density matrices of one of the parties \cite{Abad}. The most recent one is based on the local quantum uncertainty for observables of one part, the uncertainty being related to the correlations in a bi-partite state. This measure is denoted by $LQU$ and it has a closed form for $2$ by $d$ dimensional systems \cite{LQU}. 
Note that the geometric measure of correlation \cite{GeometricDiscord} can easily be computed for the states (\ref{rhouu}) and (\ref{rhoud}), in fact regarding to this correlation measure both these states have the same amount of correlations. Nevertheless we do not use this correlation measure since it has the undesirable property of increasing under local reversible operations of part $B$ \cite{Piani}. In this section we consider $LQU$ to compare the correlation content of $\rho^{\u\u}$ and $\rho^{\u\d}$. 
For our discussion it is also important to pay attention to a discrete measure of correlation, called rank \cite{Gessner} which shows how much useful a state is for a specific quantum processing task \cite{Wang}, we will discuss about this measure in section (\ref{preparation}). (Note that a quantum correlation measure need not to be symmetric with respect to two parties, here we calculate the correlations with regard to part A).\\

%
%
%
%

The definition of Local Quantum Uncertainty (LQU) is based on the observation that the existence of correlation with a far away party $B$, prevents exact determination of even a single observable in a state possessed by a party $A$. It is defined as \cite{LQU}

\be\label{lqudef}
\mathcal{U}_{A}(\rho)=\min_{K_{A}} \mathcal{I}(\rho,K_{A} \otimes I_B),
\ee
in which $K_A$ is an observable on part $A$ and
\be
\mathcal{I}(\rho,K)=-\frac{1}{2} Tr\{[\rho^{\frac{1}{2}},K]^{2}\},
\ee
is called the skew information \cite{skew}.
For a $2\times d$ dimensional system, (\ref{lqudef}) can be cast into a closed form and is given by \cite{LQU}

\be
\mathcal{U}_{A}(\rho)=1-\lambda_{max}\{W\},
\ee
where $\lambda_{max}$ denotes the largest eigenvalue, and $W$ is a symmetric matrix with elements
\be
W_{ij}=tr\{\rho^{\frac{1}{2}} (\sigma_{i} \otimes I) \rho^{\frac{1}{2}} (\sigma_{j} \otimes I)\},
\ee
with $i,j=1,2,3$.
\\

For the state (\ref{rhouu}),  the square root of the density matrix can be
calculated in closed form. Lengthy but straightforward
calculation gives the local quantum uncertainty in terms of the
eigenvalues of the matrix $\rho^{\u\u}$, which are \ba
\lambda_{0}=\frac{1}{4}(1-t_{1}-t_{2}-t_{3}),\cr
\lambda_{1}=\frac{1}{4}(1-t_{1}+t_{2}+t_{3}),\cr
\lambda_{2}=\frac{1}{4}(1+t_{1}-t_{2}+t_{3}),\cr
\lambda_{3}=\frac{1}{4}(1+t_{1}+t_{2}-t_{3}). \ea The final
result is 
\be\label{LQUuu}
LQU(\rho^{\u\u})=1-\max_{i}{\{w_{i}\}},
\ee 
in which 
\ba\label{Wi}
w_{i}&=&2(\sqrt{\lambda_{0}
\lambda_{i}}+\sqrt{\lambda_{i+1} \lambda_{i+2}})\cr &=&
\frac{1}{2}\left(\sqrt{(1-t_i)^2-(t_{i+1}+t_{i+2})^2}+\sqrt{(1+t_i)^2-(t_{i+1}-t_{i+2})^2}\right),\ea
and the summations in the subscripts are done in mod 3.
Correspondingly for $\rho^{\u\d}$, we use the same formula as in
(\ref{Wi}), with all $t_i$ replaced with $-t_i$. The important point now is that for the general case this measure is not symmetric under the change $t_i \leftrightarrow -t_i$, but when one of the $t_i$'s is equal to zero LQU will be the same for both types of the states $\rho^{\u\u}$ and $\rho^{\u\d}$. To see this explicitly, let us fix one of the $t_i$'s say
$t_3=0$. Then we find from (\ref{Wi}) that
\ba
w_1&=&\frac{1}{2}(\sqrt{(1-t_1)^2-t_2^2}+\sqrt{(1+t_1)^2-t_2^2}),\cr
w_2&=&\frac{1}{2}(\sqrt{(1-t_2)^2-t_1^2}+\sqrt{(1+t_2)^2-t_1^2}),\cr
w_3&=&\frac{1}{2}(\sqrt{1-(t_1+t_2)^2}+\sqrt{1-(t_1-t_2)^2}), \ea
which in view of (\ref{LQUuu}),  clearly shows the symmetry
$LQU(t_1,t_2)=LQU(-t_1,-t_2)$. This is to be expected owing to 
the fact that when $t_3=0$, a local transformation ($\sigma_3\otimes I$),
turns $\rho^{\u\u}$ in (\ref{rhouu}) into $\rho^{\u\d}$ in (\ref{rhoud}).\\

For the general case to see the difference of LQUs quantitatively, let us use (\ref{Wi}) and compare $w_i$  for
$\rho^{\u\u}$ and $\rho^{\u\d}$. A simple calculation shows that 
\be
w_i^2(\rho^{\u\u})-w_i^2(\rho^{\u\d})=\frac{1}{2}\left(\sqrt{a-8t_1t_2t_3}-\sqrt{a+8t_1t_2t_3}\right),
\h \forall \  i, \ee where 
\be\label{def a}
a=(1-t_1^2-t_2^2-t_3^2)^2-4(t_1^2t_2^2+t_1^2t_3^2+t_2^2t_3^2).\ee
This obviously confirms our earlier result that when at least one of the $t_i$'s is equal to zero $4$ (i.e. $t_1t_2t_3=0$), the quantum correlations for $\rho^{\u\u}$ and $\rho^{\u\d}$ are equal and further shows that when $t_1t_2t_3>0$,
 then for all $i$, $w_i(\rho^{\u\u})<w_i(\rho^{\u\d})$ and hence $$LQU(\rho^{\u\u}) > LQU(\rho^{\u\d}).$$
 It is seen that the larger is the parameter $t_1 t_2 t_3$, the higher is the difference between the quantum correlations.\\
 
\begin{figure}[tp]
\begin{center}
\vskip -5mm
\includegraphics[width=10cm, height=7.1cm]{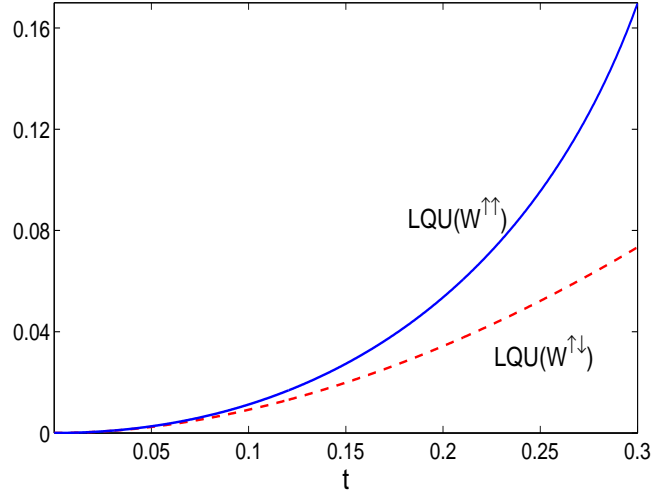}
\caption{(Color online)   The quantum correlation of the parallel and anti-parallel Werner states as  compared by Local Quantum Uncertainty (LQU). The solid line shows LQU for 
$W^{\u\u}(t)$ and the dashed line shows it for  $W^{\u\d}(t)$ . Both LQU and t are dimensionless quantities.} \label{LQUfig}
\end{center}
\end{figure}

As a simple but important special case, we look into the Werner states (\ref{WernerAll}), by putting $t_1=t_2=t_3=-t$. In this case, one finds from (\ref{Wi}) that 
\be
w_1=w_2=w_3=\frac{1}{2}(\sqrt{(1+3t)(1-t)}+1-t),\ee 
from which we can find the LQU of both the states $W^{\u\u}(t)$ and $W^{\u\d}(t)$,
depicted in figure (\ref{LQUfig}), where it is clearly seen that the LQU of $W^{\u\u}(t)$ is higher than that of $W^{\u\d}(t)$ and the difference becomes maximum when $t=1/3$ in which case the two states become uniform mixtures of parallel and anti-parallel spin states in the three directions $x,y$ and $z$.

\section{Preparation method of separable two qubit states}\label{preparation}
As we stated in the introduction the problem of preparation of entangled states (as quantum resources) has been investigated. Since there is growing evidence that separable states which have some degree of quantum correlation can be useful for quantum information and communication tasks \cite{Wang, RSP}, here we study the problem of preparation of separable quantum correlated states.
In contrary with entanglement, other quantum correlations can be generated by applying quantum channels that act \textit{only} on one of the parties of the state \cite{Bruss}. Consider the separable two qubit state $\rho_{cc}=\frac{1}{2}\left(|00\ra\la00|+|11\ra\la 11| \right)$  which is classically correlated and has zero quantum correlation since it is diagonal in the tensor product computational basis of two qubits. If the channel $\mathcal{E}$ with Kraus operators $E_1=|0 \ra\la 0|$ and $E_2=|+ \ra\la 1|$ acts on the first party of $\rho_{cc}$, the final state will be $\rho_{qc}=\frac{1}{2}\left(|00\ra\la00|+|+1\ra\la +1| \right)$ which can not be diagonalized in any tensor product basis of two qubits and hence it has non-zero quantum correlations (with regard to part A).

It is worthwhile to note that while some quantum correlated states can be produced by local operations, there are indeed some quantum states like $W^{\u\u}$ and $W^{\u\d}$ ((\ref{W'}) and (\ref{W})) which can not be created by local operations on any classical state. 
In fact there is a discrete correlation measure which can identify local producibility of quantum correlations, it is the rank $R$ of the correlation matrix of the state \cite{Gessner}. $R$ is nothing but the
number of orthogonal operators which is needed in the expansion of a density matrix. In fact rank cannot be increased by local operations of one party alone \cite{Gessner}. Since all classically correlated states are of rank 2, applying local operations on such states can produce only rank-2 quantum correlated states like $\rho_{qc}$ \cite{Bruss}. On the other hand it is certain that no rank-3 or rank-4 state can be produced by local operations of one of the parties on a classically correlated state \cite{Gessner} and hence these states certainly have not locally producible quantum correlations.\\

It is also worthwhile to note that states with different correlation ranks act in a different manner in quantum information tasks and it seems that states of higher ranks, which can not be created locally, are more useful for such tasks. For example while maximally entangled states are used for teleportation, separable states of rank-4 can be used for sending the information required for reconstruction of an arbitrary state by a remote party.  For rank-3 states, the method is used to reconstruct only pure states \cite{Wang}. Regarding the effectiveness of not locally producible quantum correlated states in quantum information tasks, it will be an important question that how one can prepare these states and what kind of operations is necessary for the preparation?\\

Note that the correlation rank of the states $\rho^{\u\u}$ and $\rho^{\u\d}$ are both equal and that is $1+$ number of non-zero $t_i$s. Here we present the preparation method of all these states with all correlation ranks.\\

To present the preparation method we show that one can start from a simple classically correlated state
\be\label{rhoCCuu}
\rho^{\u\u}_{cc}=\frac{1}{2}(|00\ra\la 00|+|11\ra\la 11|),
\ee
and produce all the states in the class $\Lambda^{\u\u}$ by bi-local unitary actions. It has been shown in \cite{generation} how one can experimentally prepare such classically correlated states in the valence electrons of two $^{40}Ca+$ ions in a linear Paul trap, where a qubit is encoded in an $S^{1/2}$ ground and a $D^{5/2}$ metastable state. Once this state is prepared,  Alice and Bob act on their initial state (\ref{rhoCCuu}) by the correlated unitary channel
 \be
 {\cal E}(\rho)=\sum_{i=0}^3 p_i(U_i\otimes V_i)\rho (U_i^{\dagger}\otimes V_i^{\dagger}),
 \ee
 where $\{p_i\}$ is a probability distribution, $U_i$ and $V_i$ are unitary operators on single qubits and $U_0=V_0=I$ . 
 We explicitly show that by a specific choice of $U_i$ and $V_i$ one can produce all the states of $\Lambda^{\u \u}$ only by appropriate choice of $p_i$.
The same type of production is possible for the states in $\Lambda^{\u \d}$  if we start from the following classically correlated state,
\be\label{rhoCCud}
\rho^{\u\d}_{cc}=\frac{1}{2}(|01\ra\la 01|+|10\ra\la 10|).
\ee

To see this, let us define two types of Hadamard operators $H=\frac{1}{\sqrt{2}}\left(\begin{array}{cc} 1 & -1 \\ 1 & 1\end{array}\right)$ and
  $K=\frac{1}{\sqrt{2}}\left(\begin{array}{cc} 1 & i \\ i & 1\end{array}\right)$. The operator $H$ turns the basis states $|0\ra $  and $|1\ra$ into $|+\ra$ and $|-\ra$ respectively. Similarly the operator $K$ turns these states into $|y_+\ra$ and $|y_-\ra$.\\

{\bf Temporary change of notation:}
For brevity, in the few lines below, we use the notations $\rho^{\u\u}\equiv \rho^+ $ and $\rho^{\u\d}\equiv\rho^-$  and later we resort to our earlier notation. \\
\\
From (\ref{rhoCCuu}) and (\ref{rhoCCud}), we can verify the following equations:
 \ba
 (H\otimes H)\rho^{\pm}_{cc}(H^\dagger\otimes H^\dagger)=\frac{1}{4}(I \otimes I \pm\sigma_1\otimes \sigma_1),\cr
 (K\otimes K)\rho^{\pm}_{cc}(K^\dagger\otimes K^\dagger)=\frac{1}{4}(I \otimes I\pm \sigma_2\otimes \sigma_2),\cr
 \rho^{\pm}_{cc}+(\sigma_1\otimes I)\rho^{\pm}_{cc}(\sigma_1\otimes I)=\frac{I\otimes I}{2}.
 \ea
 Inserting these into (\ref{rhouu}), and (\ref{rhoud}) we find\\
 \ba\label{E}
 \rho^{\pm}&=&\frac{1-t_1-t_2+t_3}{2}\rho^{\pm}_{cc} + \frac{1-t_1-t_2-t_3}{2}(\sigma_{1}\otimes I)\rho^{\pm}_{cc}(\sigma_1\otimes I)\cr\cr
 &+&t_1 H\otimes H \rho^{\pm}_{cc} H^\dagger\otimes H^\dagger+ t_2 K\otimes K \rho^{\pm}_{cc} K^\dagger\otimes K^\dagger=:{\cal E}(\rho^{\pm}_{cc}).
 \ea
This relation defines the bi-local channel ${\cal E}$ which produces $\rho^{\u\u}$ from $\rho^{\u\u}_{cc}$ and also $\rho^{\u\d}$ from $\rho^{\u\d}_{cc}$ (see figures (\ref{BasicChannel}) and (\ref{ParalleAntiParallel})).
In view of the fact that $\sigma_1=-i K^2$, we have shown that starting from the classical states (\ref{rhoCCuu}) or (\ref{rhoCCud}), Alice and Bob should only use two types of unitary gates (rotations) to produce any state in the classes $\Lambda^{\u\u}$ or $\Lambda^{\u\d}$ and by appropriate choice of the parameters $t_i$, they can prepare states of all ranks.\\

 This method of preparation is more efficient than the recently proposed one \cite{generation}, that method is used to produce \textit{some} rank-4 or rank-3 states by applying a \textit{continuous spectrum} of bi-local rotations of the form $\frac{1}{2\pi} \int_0^{2\pi} R_{\vec{n}}(\theta) \otimes R_{\vec{n}}(\theta) \rho R_{\vec{n}}(\theta)^\dagger \otimes R_{\vec{n}}(\theta)^\dagger$, with $R_{\vec{n}}(\theta)=e^{-i\theta \vec{n}.\vec{\sigma}/2}$,  on the classical state $\rho^{\u\u}_{cc}$. 
Hence to compare our method with that of \cite{generation}, we stress on the fact that using our method one can prepare \textit{all} separable two qubit states of \textit{all} ranks by applying only \textit{two} types of rotations $H$ and $K$. Again note that the initially used classical states (\ref{rhoCCuu}) and (\ref{rhoCCud}) can experimentally be prepared in the valence electrons of two $^{40}Ca+$ ions in a linear Paul trap. For such qubit states the rotations can be realized by applying magnetic fields, hence once the states (\ref{rhoCCuu}) and (\ref{rhoCCud}) were prepared one can use the same setup as the one proposed in \cite{generation} to apply proper magnetic fields in order to implement the channel $\mathcal{E}$ of equation (\ref{E}).\\

In the next section we show that this preparation method can also explain the origin of the correlation difference of the states $\rho^{\u\u}$ and $\rho^{\u\d}$, in fact we relate this difference to the extra amount of hidden classical correlation which is required in the preparation of $\rho^{\u\u}$.

\section{Origin of correlation difference}\label{origin}

\begin{figure}[tp]
\begin{center}
\vskip -5mm
\includegraphics[width=14cm, height=5cm]{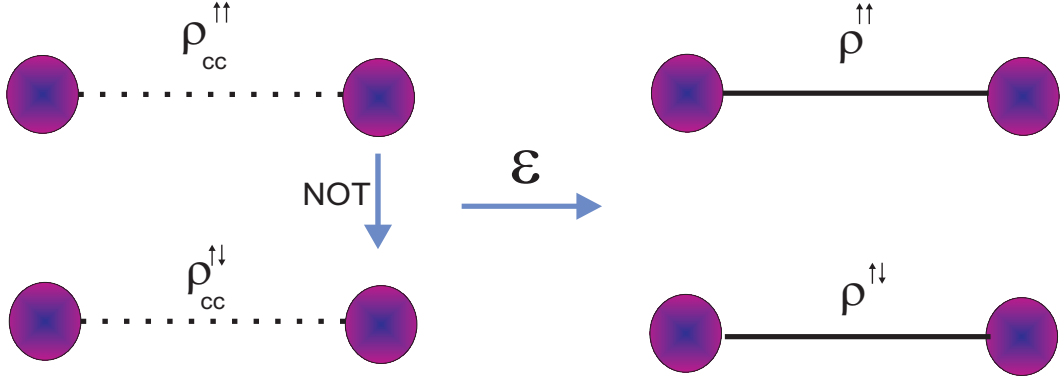}
\vskip 1mm
\caption{(Color online)  The states ${\rho_{cc}}^{\u\u}$ and ${\rho_{cc}}^{\u\d}$ have the same amount of classical correlation, as they can be converted to each other by a local NOT operator. However the resulting states $\rho^{\u\d}$ and $\rho^{\u\u}$,  cannot be locally converted to each other and do have different amounts of quantum correlations, despite the fact that both of them have been produced by the same channel.}
\label{ParalleAntiParallel}
\end{center}
\end{figure}

In this section we want to see on physical and operational grounds, why the correlations of the two states $\rho^{\u\u}$ and $\rho^{\u\d}$ are different. 
To see more clearly the relevance of this question, note that the difference of the two initial classically correlated states (\ref{rhoCCuu}) and (\ref{rhoCCud}), used for production of states in $\Lambda^{\u\u}$ and
 $\Lambda^{\u\d}$,  is just a simple local unitary rotation
  $|0\ra\leftrightarrow |1\ra$. Such a local action does not produce any quantum correlation. 
  Nevertheless when this simple unitary action is followed by the channel ${\cal E}$ (defined in (\ref{E})), it leads to two
   states $\rho^{\u\u}$ and $\rho^{\u\d}$ with manifestly different quantum correlations (see Fig. (\ref{ParalleAntiParallel})). The question is why  this simple local action on the initial state, produces
   different quantum correlation at the end?  What is the source of
   this excess quantum correlation, despite our equal action on the two states $\rho^{\u\u}_{cc}$ and $\rho^{\u\d}_{cc}$?
    Is there any kind of hidden classical correlation in the initial state or in the channel which is converted into the
    final quantum correlation in the resulting states and makes their correlation different? \\

    In the following we will argue
    that there is indeed a hidden classical correlation in the channel which causes this discrepancy. In fact, the difference of the states in $\Lambda^{\u\u}$ and
 $\Lambda^{\u\d}$ should be traced back to the requirement of setting up a standard and agreed-upon frame of coordinate axes between Alice and Bob.
We will show that to enact the channel ${\cal E}$ on the initial
state (\ref{rhoCCuu}) and produce the state $\rho^{\u\u}$ with a given
fidelity, they need to align their coordinate axes with more
precision compared with the case when they want to produce the
state $\rho^{\u\d}$ from (\ref{rhoCCud}). This more precise alignment costs them
sending back and forth a larger number of bits before they start
the process. It is this extra communication of bits which goes
into the final higher value of quantum correlation for the states
$\rho^{\u\u}$. Note that in the present work,  our method of comparison is not entirely
quantitative in this respect, that is, while we prove the above
statements quantitatively, we do not exactly relate the
extra quantum correlation obtained to the extra classical
correlation which is necessary for aligning the coordinate
systems. Our argument runs as follows.\\

 First we should note that the two parties agree on the $z$ axis, since it is assumed that they both agree on the form of classically correlated states which has been given to them.  What they need to do is to agree on the coordinate axes $x$ and $y$ to enact on this state by their local rotation operators:

\be
H=R_{y}(\frac{\pi}{2}),\h K=R_x(-\frac{\pi}{2}), \h \sigma_1=-i R_x(\pi).
\ee
Note that the operator $\sigma_1$ is applied only by Alice. This is nothing but a $\pi$ rotation around any axis in the $x-y$ plane (which is perpendicular to the $z$ axis
  and hence is known to both Alice and Bob).  The difficulty arises when they want to enact in a correlated way
  the unitary operations $R_y(\frac{\pi}{2})$ or $R_x(-\frac{\pi}{2})$ for which they have to agree on a fixed axis in the $x-y$ plane (For example by using the method proposed in \cite{Gisin}, or more generally by using the well know literature about the Reference Frames \cite{RF-review, RF1}). Once this axis is chosen the other axis is automatically chosen to be perpendicular to this one and lying in the $x-y$ plane. It should be noted that here we investigate the effect of the alignment of the axes in the preparation process, and the method which Alice and Bob have used to align their axes is not the matter of interest.\\  
  
\begin{figure}[tp]
\begin{center}
\vskip -5mm
\includegraphics[width=14cm, height=6cm]{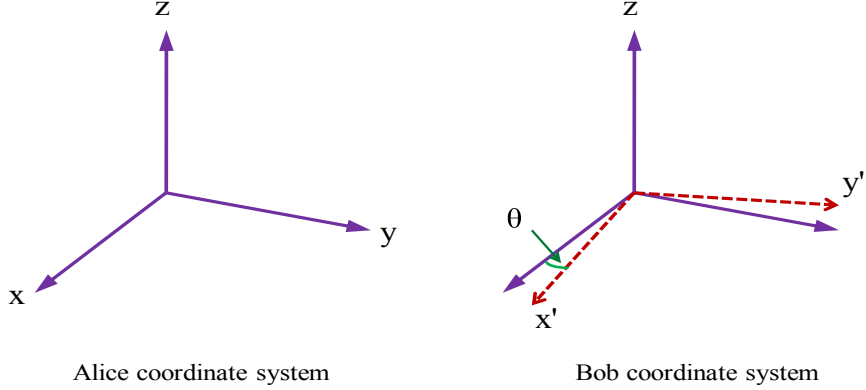}
\vskip -5mm
\caption{(color online) To turn the classically correlated states $\rho_{cc}$ to quantum correlated states, Alice and Bob need to correlate their unitary actions on their qubits. This requires  precisely aligned coordinate axes between them. For producing $\rho^{\u\u}$ they need more precise alignment. This extra correlation in setting up the axes and enacting of correlated channels goes into the final quantum correlation of the $\rho^{\u\u}$ state.  }
\label{coordinate}
\end{center}
\end{figure}

Suppose now that Alice sets up an $x$ axis in the $x-y$ plane while Bob's $x$ axis is not exactly aligned with the $x$ axis of Alice, but is rotated with respect to it by an angle $\theta$ in the $x-y$ plane. Hence instead of $x$ and $y$ axes, Bob has considered $x'$ and $y'$ (see Fig. \ref{coordinate}).
The correlated channel which now Alice and Bob enact on the states $\rho_{cc}^\pm$ is denoted by ${\cal E}_{\theta}$ rather than ${\cal E}$, where

 \ba
{\cal E}_{\theta}(\rho^{\pm}_{cc})&=&\frac{1-t_1-t_2+t_3}{2}\rho^{\pm}_{cc} + \frac{1-t_1-t_2-t_3}{2}(\sigma_{1}\otimes I)\rho^{\pm}_{cc}(\sigma_1\otimes I)\cr\cr
 &+&t_1 H\otimes H' \rho^{\pm}_{cc} H^\dagger\otimes H'^\dagger+ t_2 K\otimes K' \rho^{\pm}_{cc} K^\dagger\otimes K'^\dagger,
 \ea
 in which $H'=R_{y'}(\frac{\pi}{2})$ and $K'=R_{x'}(-\frac{\pi}{2})$. The state which is prepared in this way differs from what they wanted to prepare. In fact we see that
\be \label{E'}
{\cal E}_{\theta}(\rho^{\pm}_{cc})=(I\otimes R_z(\theta)) {\cal E}(\rho^{\pm}_{cc})(I\otimes R^{\dagger}_z(\theta)),
\ee
where $\theta$ is the angle of $x$ axis of Bob with respect to that of Alice.
Let us see how much this error in aligning the $x$ axis affects the final state.
We measure this by the fidelity of the resulting state and the
desired state. In the appendix it is shown how the fidelity can
be calculated. The result is as follows:

\be \label{Fidelity}\nonumber
F_{\theta}^{\u\u}:=F\left({\cal E}({\rho_{cc}}^{\u\u}),{\cal E}_{\theta}({\rho_{cc}}^{\u\u})
\right)=\frac{1}{2}\sqrt{(1+t_3)^2-(t_1-t_2)^2\sin^2
\theta}+\frac{1}{2}\sqrt{(1-t_3)^2-(t_1+t_2)^2\sin^2 \theta}. \ee
By changing $t_i$ to $-t_i$ we obtain
 \be\nonumber
F_{\theta}^{\u\d}:=F\left({\cal E}({\rho_{cc}}^{\u\d}),{\cal E}_{\theta}({\rho_{cc}}^{\u\d})
\right)=\frac{1}{2}\sqrt{(1-t_3)^2-(t_1-t_2)^2\sin^2
\theta}+\frac{1}{2}\sqrt{(1+t_3)^2-(t_1+t_2)^2\sin^2 \theta}. \ee
In order to compare the fidelities, we  simplify $
(F_{\theta}^{\u\u})^2-(F_\theta^{\u\d})^2$ and after some
rearrangements we find that:

\be
(F_\theta^{\u\u})^2-(F_\theta^{\u\d})^2=\frac{1}{2}\left(\sqrt{a_{\theta}-8t_1t_2t_3\sin^2\theta}-\sqrt{a_{\theta}+8t_1t_2t_3\sin^2\theta}\right),
\ee
where $a(\theta)$ has the same form as in (\ref{def a}) except that $t_1 $ and $t_2$ should be replaced with $t_1\sin\theta$ and $t_2\sin\theta$ respectively.  This clearly shows that $(F_\theta^{\u\u})^2 < (F_\theta^{\u\d})^2$
 as long as $t_1t_2t_3>0$ and they are equal only when $t_1t_2t_3=0$. Thus we see that the difference in fidelity is larger the more distant the states are from lower rank states, as measured by the parameter $t_1t_2t_3$.\\
%
%
%

Briefly, in this section we have shown that for rank-4 states, it is always harder to prepare the state $\rho^{\u\u}$ than the state $\rho^{\u\d}$,
 in the sense that Alice and Bob need to precisely agree on their corresponding coordinate axes, otherwise they end up with a state which
 has a  lower fidelity with the required state.  When $t_1t_2t_3=0$, (i.e. when the rank is less than 4), the two fidelities are equal and
  at the same time the corresponding quantum correlations are also equal as they should be, since in this case the two states are convertible
    to each other via local unitary actions.\\

\section{An alternative method of preparation, {\large{using optimal NOT operation}}} \label{optNotSection}
One may wonder that the extra resource which is required for the preparation of $\rho^{\u\u}$ is a consequence of our preparation method, in this section we show that even using an alternative method, one needs more initial resources to prepare $\rho^{\u\u}$ compared with $\rho^{\u\d}$.
Up to now we have emphasized the absence of a universal NOT operator in our arguments (see section (\ref{preliminary}) and figure (\ref{ParalleAntiParallel})). 
As a matter of fact the universal NOT does not exist since it is not a completely positive map and it can be regarded as an anti-unitary operator.
It is well known that although a universal NOT operator does not exist, there is an optimal NOT operation which can approximate it to any desired degree of accuracy \cite{optNOT1,optNOT2,optNOT3}. Therefore it is in order to re-evaluate our arguments in the light of this finding.  This is what we do in this section. \\

In \cite{optNOT1} it is shown that, although a universal NOT channel violates quantum mechanics and hence cannot exist,  one can achieve an optimal NOT which out of $N$ copies of a single pure qubit state, produces a qubit state which can be as orthogonal as we wish to the original state.
There are two different scenarios to design the optimal NOT operator, the first method relies on optimal state estimation of the original state and then re-preparation of the orthogonal state whilst the second method would be to approximate an anti-unitary transformation on the Hilbert space of the input qubit(s) by a unitary transformation on a larger Hilbert space which describes the input qubit(s) and ancillas. The fidelity of the produced state with the actual orthogonal state is the same in  both scenarios and is given by $F=\frac{N+1}{N+2}$ \cite{optNOT1}.
It should also be noted that regarding the second approach, an experimental realization of the optimal NOT has been carried out, where stimulated emission in parametric down conversion has been used \cite{optNOT2, optNOT3}. In the experiment, one qubit is encoded in a polarization state of a single photon which is injected as the input state into an optical parametric amplifier excited by a pulsed, mode-locked ultraviolet laser beam \cite{optNOT2}.\\

The resulting process of optimal NOT operations
 can be described by the simple quantum channel \cite{optNOT1}
\be \label{opt-NOT}
\sigma^{\otimes N} \lo \Phi(\sigma^{\otimes N})=\frac{N}{N+2}\sigma^\perp+\frac{1}{N+2} I,
\ee
where $\sigma^\perp$ is the state which is orthogonal to $\sigma$. \\

\begin{figure}[tp]
\begin{center}
\vskip -5mm
\includegraphics[width=15cm, height=4cm]{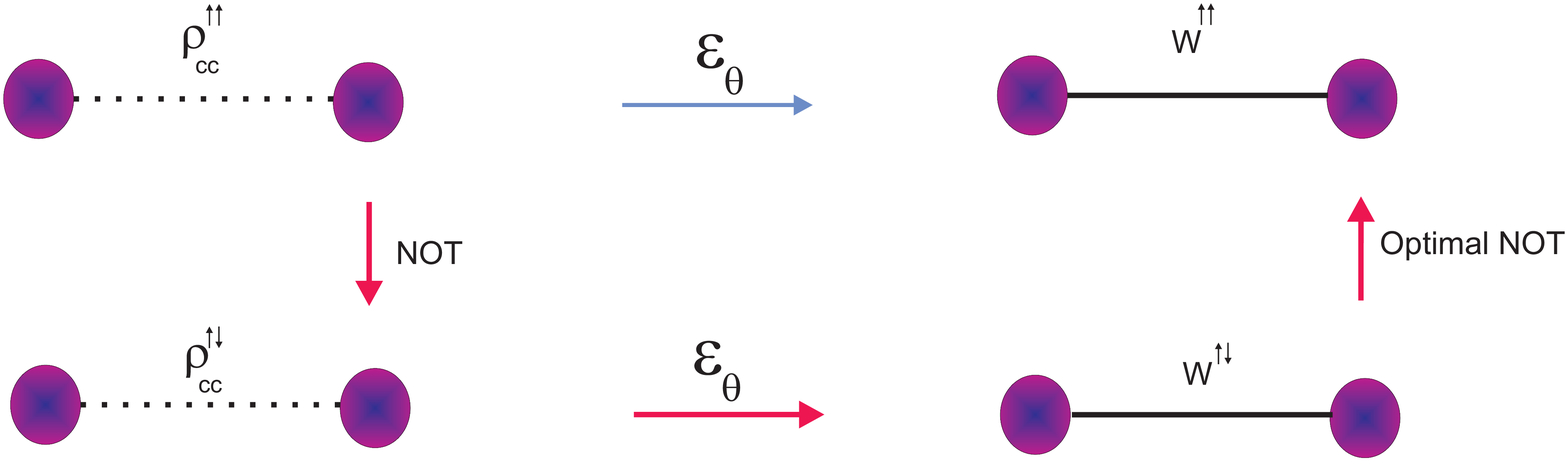}
\caption{(Color online)  Starting from a classically correlated state of parallel spins, there are two ways to produce the state $W^{\u\u}$. In method A, the channel ${\cal E}_{\theta}$ directly acts on the state ${\rho_{cc}}^{\u\u}$, in method B, first a NOT operation on one of the spins turn the state to ${\rho}_{cc}^{\u\d}$ on which the  same  channel ${\cal E}_{\theta}$ acts. Finally the optimal NOT  produces a state which is to be as close as possible to $W^{\u\u}$. Figure (\ref{resource}) compares the fidelities of the two methods. }
\label{OptimalNot}
\end{center}
\end{figure}

Equipped with this new operation, one can now imagine an alternative method for production of $\rho^{\u\u}$ states which at first sight may use less resources than the one mentioned above. For concreteness we restrict ourselves to the production of the state $W^{\u\u}(t)$. In this alternative method, one acts on the ${\rho_{cc}}^{\u\d}$ from the beginning and prepares $W^{\u\d}(t)$ without the need for much precise alignment between the axes, and then only at the end uses the optimal NOT operation (by Bob for example) to turn $W^{\u\d}(t)$ to a state as close as possible to $W^{\u\u}(t)$. This is shown in figure (\ref{OptimalNot}) along with the original method mentioned in section (\ref{preparation}). The price that one should pay is to use more copies of the initial states in order to achieve a given fidelity as per equation (\ref{opt-NOT}).  In other words, in this new method one compromises the precision in $\theta$ (the precision in alignment) for the number of pairs of states to begin with. To make a comparison between the two methods, we assume that the axes have been aligned with a precision $\theta$ and then compare the fidelities of the two methods as follows. \\

Method A) In this method, the channel ${\cal E}_{\theta}$ is applied to the classical state ${\rho_{cc}}^{\u\u}$ and we calculate the fidelity of the resulting state with an ideal $W^{\u\u}(t)$ state, which turns out from (\ref{Fidelity}) to be:

\be \label{fidmethod1}
F_A\equiv F({\cal E}_{\theta}({\rho_{cc}}^{\u\u}), W^{\u\u}(t))=\frac{1}{2}\left[1+t+\sqrt{(1-t)^2-4t^2\sin^2 \theta}\right].
 \ee

\begin{figure}[h]
\begin{center}
\vskip -5mm
\includegraphics[width=11.8cm, height=9.45cm]{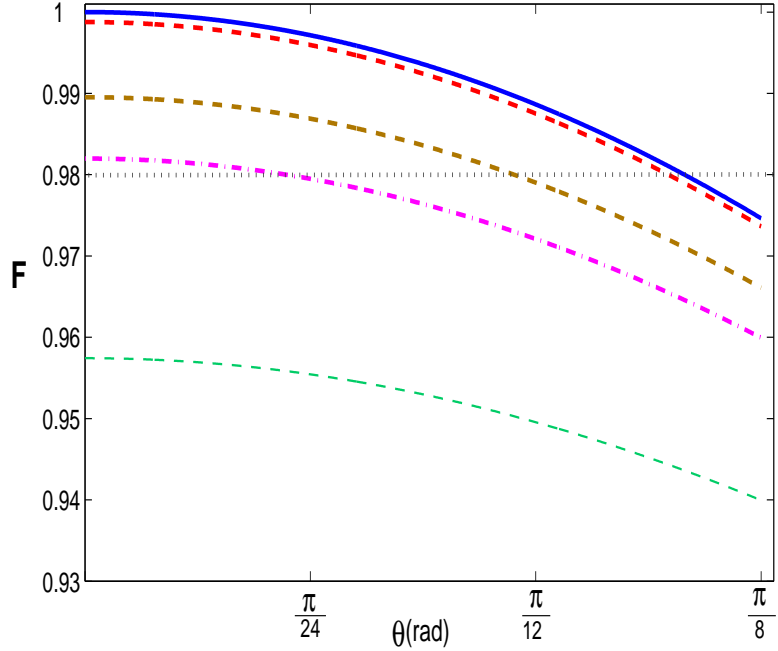}
\vskip -2mm
\caption{(Color online) Fidelities of preparing the state $W^{\u\u}$ by methods A and B.  The solid (blue) line corresponds to method A, where other curves show fidelities for different number of initial copies ($N=1,5,10$ and $100$ from bottom to top), used for the optimal NOT gate. Two features are evident: First, for any value of $\theta$ (precision in the alignments), method A gives a higher fidelity than method B, no matter how many copies are used. The two methods give the same fidelity only when the number of copies in method B is infinite. Second, to achieve a given fidelity, the two parties have two options: they can either align their axes with low precision (high $\theta$) and then use a larger number of copies for the optimal NOT process, or else, they can use precisely aligned axes, in which case they need use a lower number of copies for the optimal NOT process. All the fidelities are dimensionless quantities. } 
\label{resource}
\end{center}
\end{figure}

Method B) In this method, the channel ${\cal E}_{\theta}$ is applied to each of the N copies of the classical state ${\rho_{cc}}^{\u\d}$ and then the optimal NOT channel $(I_N \otimes \Phi)$  is applied to 
the resulting state to produce a state which is as close as possible to $W^{\u\u}(t)$. The fidelity of the resulting state is:

\ba\nonumber
F_B&\equiv& F((I_N \otimes \Phi) \bigl({\cal E}_{\theta}({\rho_{cc}}^{\u\d})\bigr)^{\otimes N} , W^{\u\u})\cr &=&\frac{1}{2\sqrt{2}}\left[2(1+t)(1+st)+\sqrt{(1-t)(1-st)+4st^2 \cos 2\theta + \sqrt{[(1-t)^2-4t^2][(1-st)^2-4s^2t^2]}}\right],
\ea
where $s=\frac{N}{N+2}$.
The two fidelities are plotted in figure (\ref{resource}) for $t=\frac{1}{3}$ which corresponds to the state $W^{\u\u} \equiv W^{\u\u}(\frac{1}{3})$.  The results shown in this figure, confirm our earlier result that some sort of hidden classical correlation in the channel is responsible for the higher quantum correlation in the final state $W^{\u\u}$. This hidden classical correlation shows itself in method A as the need for more precise alignment of axes, and in method B as using a larger number of classically correlated states to begin with in order to achieve the final fidelity with the desired states.
 In both cases Alice and Bob need a larger amount of classical correlation for producing the state $W^{\u\u}$.\\

Having established the difference of quantum correlations between the states of $\Lambda^{\u\u}$ and $\Lambda^{\u\d}$, we can now ask whether the higher amount of quantum correlation makes the parallel states more suitable for quantum communication tasks. This is indeed the case as we show in the next section for one example. Clearly this is only an example and does not exhausts all the communication tasks.

\section{Parallel and anti-parallel Werner states in application}\label{effect}

As we mentioned before, separable states which have some degree of quantum correlation can be useful for quantum communication tasks like transmission of the information of the states, in which the rank of the correlation matrix determines the usefulness of the state \cite{Wang}. 
 Other examples include Remote State Preparations (RSP) \cite{RSP} where the fidelity is related to the geometric measure of correlation inherent in the shared state between the two parties \cite{discordRSP}. Therefore one cannot compare the effectiveness of the parallel and anti-parallel states $\rho^{\u\u}$ and $\rho^{\u\d}$ by such tasks. To unravel a difference we resort to another task, namely distribution of entanglement by using separable parallel and anti-parallel Werner states.\\

When a resource is used for quantum information tasks, a natural question is that how much robust that resource is against the noise. We know that entanglement is a fragile quantum resource, i. e. when an entangled system is exposed to noise, the entanglement starts leaking. This observation leads us to the intuition that the states which are more quantum correlated, are more fragile when exposed to a noise. We will see in this section, this is indeed the case for the states $\rho^{\u\u}$ and $\rho^{\u\d}$.

\subsection{Entanglement sharing with Werner states}

In the protocols of entanglement distribution with separable states \cite{Cubitt, Peuntinger}, a separable state $\rho_{AB}$ is shared between two parties, Alice and Bob. Alice adds a qubit $C$, performs a local operation and sends it to Bob who after a local operation again, will change the original separable state into a mixed state with a definite and non-zero entanglement. The original separable state $\rho_{AB}$ is of a special kind and  the mediated qubit $C$ remains separable with the states of $A$, $B$ and $AB$ throughout the process \cite{Cubitt}. We now want to see which one of the Werner states $W^{\u\u}(t)$ or $W^{\u\d}(t)$ are more effective for the above task. Given the above results, one expects that  $W^{\u\u}(t)$ may lead to more quantum entanglement than $W^{\u\d}(t)$ in such a protocol. As we will see, this is truly the case.\\

Suppose that the state $W^{\u\u}(t)$ is shared between Alice and Bob. Alice prepares the ancilla $C$ in the initial state
$| + \ra$ in her possession and  uses her particle $A$ as a controller to apply a $Z$ gate on the ancillary particle $C$, then she sends $C$ to Bob who performs another $CZ$ gate on the particles $B$ and $C$.
The initial state of the particles $ABC$ can be written as
\begin{equation}
\rho^{0}_{ABC}=W^{\u\u}_{AB}(t) \otimes | + \ra\la + |_{C},
\end{equation}
and the final state after the operations of Alice and Bob is
\begin{equation}\label{final}
\rho_{ABC}=\frac{1+t}{2} \rho^{+}_{AB} \otimes |+\ra\la+|_{C} + \frac{1-t}{2} \rho^{-}_{AB}(t) \otimes |-\ra\la -|_{C}.
\end{equation}

\begin{figure}[tp]
\begin{center}
\vskip -5mm
\includegraphics[width=10cm, height=7.5cm]{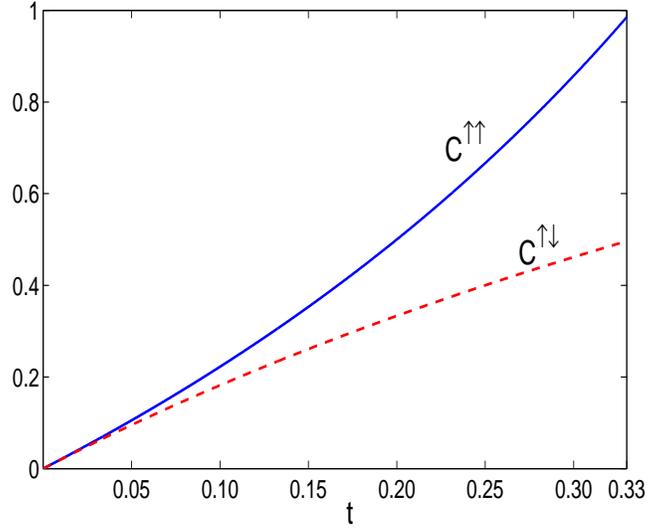}
\vskip -2mm
\caption{(Color online)  The entanglement of the final states in the entanglement distribution protocol when the initial state is $W^{\u\u}(t)$ (solid blue line) or $W^{\u\d}(t)$ (dashed red line). Entanglement is quantified by concurrence which is dimensionless and t also reperents the dimensionless parametter of eq (\ref{concfinal}). }
\label{conc}
\end{center}
\end{figure}

If  Bob measures $C$ in the  $x$ basis,  he will find $+1$ with  probability $p_{+}=\frac{1+t}{2}$ and the state of $AB$ collapses to the separable state $\rho^{+}_{AB}=\frac{1}{2}\left(\left|00\right>\left<00\right|+\left|11\right>\left<11\right|\right)$ which is a classically correlated state. He may also find $-1$ with $p_{-}=\frac{1-t}{2}$ where the state of $AB$ will change to an entangled state $\rho^{-}_{AB}$:
\begin{equation}\label{final-}
\rho^{-}_{AB}(t)=\frac{1}{4}\left( I \otimes I + \frac{2t}{1-t}  \sigma_{1} \otimes \sigma_{1} + \frac{2t}{1-t}  \sigma_{2} \otimes \sigma_{2} -  \sigma_{3} \otimes \sigma_{3}\right).
\end{equation}
One can easily change $t$ to $-t$ and get the final state of the process when the shared separable state is $W^{\u\d}(t)$.
The entanglement of the final states can be compared by calculating their concurrences \cite{Wootters}. The result is: 
\begin{eqnarray}\label{concfinal}
C^{\u\u}(t):=C(\rho^-_{AB}(t))&=&\frac{2t}{1-t} \hspace{1.5cm} 0\leq t \leq \frac{1}{3}, \nonumber\\
C^{\u\d}(t):=C(\rho^-_{AB}(-t))&=&\frac{2t}{1+t} \hspace{1.5cm} 0\leq t \leq \frac{1}{3}.
\end{eqnarray}
The result is plotted in figure (\ref{conc}), from which it is evident that the state, $W^{\u\u}(t)$ does perform better than the state $W^{\u\d}(t)$ in this process. \\

\subsection{Robustness of Werner states against noise}  

In this subsection we investigate the effect of noise on separable quantum correlated states. For simplicity and for definiteness we again consider only Werner states. (We could have considered general states of the form (\ref{rhouu}) and (\ref{rhoud}) for this purpose, but the essential feature is also revealed in this special isotropic case). 
A natural type of noise which retains the isotropy of these states is the depolarizing noise   $\rho\lo {\Phi}_p(\rho)=(1-p)\rho+p\frac{I}{2}$ 
acting on one of the qubits. It is readily found that such a noise, when
acting on the states $W^{\u\u}(t)$ and $W^{\u\d}(t)$ has the simple effect of changing $t$ to
$t'=t(1-p)$ in both cases. The new states are thus given by 
$W_p^{\u\u}(t):=W^{\u\u}((1-p)t)$ and $W_p^{\u\d}(t):=W^{\u\d}((1-p)t)$.
We can now ask two different questions, namely:\\
\begin{itemize}

\item{
What are the quantum correlations of the noisy states $W_p^{\u\u}$ and $W_p^{\u\d}$, as measured by LQU,}

and
\item{
How much the original states have been affected by noise, as measured by their fidelities with the noisy states, i.e.  $F(W^{\u\u}(t),W_p^{\u\u}(t))$ and $F(W^{\u\d}(t),W_p^{\u\d}(t))$}.
\end{itemize}

\begin{figure}[tp]
\begin{center}
\vskip -5mm
\includegraphics[width=10cm, height=7cm]{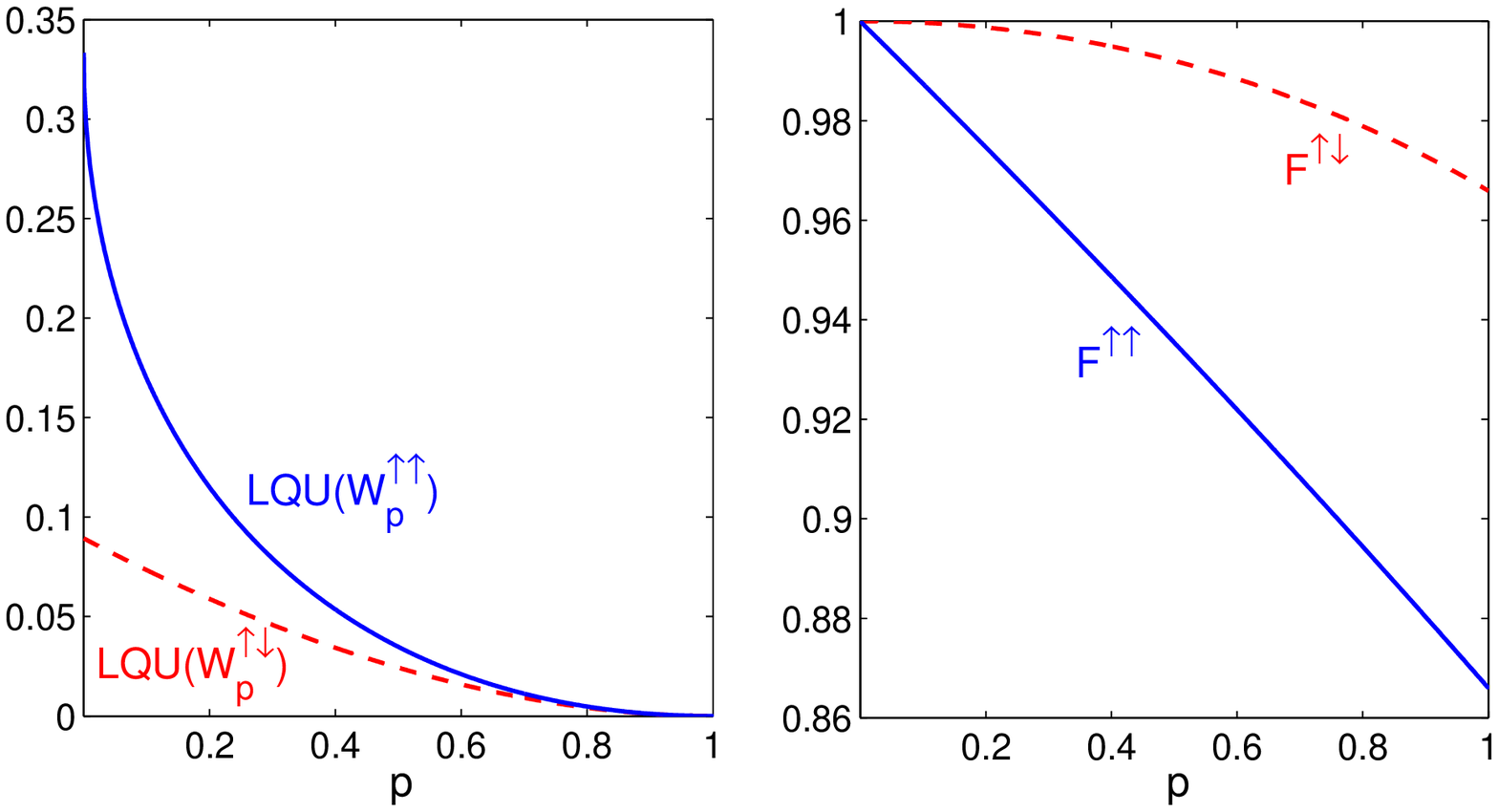}
\vskip 1mm
\caption{(Color online) Robustness of quantum correlation of parallel (solid blue lines) and anti-parallel (dashed red lines) Werner states, when both states undergo a local depolarizing channel. We have fixed the parameter $t$ to $t=\frac{1}{3}$ in both cases. Left- Quantum correlation of the deformed states as measured by local quantum uncertainty. Right- Robustness as measured by the fidelity of the original and deformed states, as given by $F^{\u\u}:= F(W^{\u\u}(\frac{1}{3}),W_p^{\u\u}(\frac{1}{3}))$ and $F^{\u\d}:=F(W^{\u\d}(\frac{1}{3}),W_p^{\u\u}(\frac{1}{3}))$. Although the deformed states have less fidelity with the original ones in the parallel case, they retain their higher value of quantum correlation. All the presented quantities are dimensionless.}
\label{FidLQU}
\end{center}
\end{figure}

The answer to the first question is readily found by using equation (\ref{LQUuu}) for LQU of isotropic states and one finds 

 \be LQU(
W_p^{\u\u}(t)=1-\frac{1}{2}\left(\sqrt{(1-3t(1-p))(1+t(1-p))}+1+t(1-p)\right).
\ee
For the other state $W^{\u\d}(t)$, it is enough to change $t$ to $-t$ everywhere in the above formula.
 Figure (\ref{FidLQU}) shows the plot of quantum correlations (LQU) for these two states for the case $t=\frac{1}{3}$ versus $p$.
It is clearly seen that the state $W^{\u\u}$ when affected by noise keeps its higher value of quantum correlation for all values of $p$ compared with $W^{\u\d}$.  We note in passing that
the same feature is also observed if both parties are subject to
depolarizing channels with parameters $p$ and $p'$, since in this case it can be easily shown that the parameter $t$ 
changes to $t(1-p)(1-p')$ and the previous argument is again valid in this case.\\
  
To answer the second question we need a closed formula for the fidelity of two general $W$ states, $W(t)$ and $W(t')$. 
Regarding the fact that $W(t)$ and $W(t')$ commute and their eigenvalues can easily be obtained (in the form of $\frac{1-t}{4} (3)$ and $\frac{1+3t}{4}(1)$, where the numbers in parentheses indicate degeneracies), we have: 
\be \label{fidWW'}
F(W(t),W(t'))=\frac{1}{4}\left(3\sqrt{(1-t)(1-t')}+\sqrt{(1+3t)(1+3t')}\right).
\ee
Using this formula we can find closed forms for the fidelities $F(W^{\u\u}(t),W_p^{\u\u}(t))$ and $F(W^{\u\d}(t),W_p^{\u\d}(t))$. For definiteness we consider the case where $t=1/3$. The result is shown in figure (\ref{FidLQU}). It is seen that the state $W^{\u\u}$ is less robust than the state $W^{\u\d}$. This is perhaps expected in view of its higher correlation. In other words, this result implies that a state which has a  higher quantum correlation is  more fragile under local noise. \\

Summing up the results of this section, we have shown that the parallel Werner states $W^{\u\u}(t)$ are more effective for entanglement sharing and yet they are more fragile against noise, which is the price one should pay for using them as a more powerful resource.

\section{Discussion}\label{discussion}

We have made a detailed study of correlation properties of all two qubit separable states with maximally mixed marginal, namely the states of the form $\rho=\frac{1}{4}\left( I \otimes I + t_i \sigma_i \otimes \sigma_i \right)$. With regard to the sign of the parameter $t_1 t_2 t_3$ these states are divided into two separate classes with representative elements $\rho^{\u\u}$ or $\rho^{\u\d}$. 
These states, while seeming very similar, have quite different correlation properties. This difference relates to a hidden classical correlation which is needed for preparation of these states, i.e. although both states are prepared by acting on equally classically correlated (parallel or anti-parallel) states by the same quantum channel, the quantum correlation in $\rho^{\u\u}$ is higher than $\rho^{\u\d}$, due to the extra hidden classical correlation required in its preparation. 
Throughout the paper we have emphasized the essential difference of these two states in that they cannot be converted to each other exactly by a universal NOT operator. In fact for lower rank states which are convertible to each other by NOT operators, such difference in correlation vanishes.
One can use an alternative method of production of parallel states from anti-parallel classically correlated states, simply by using an optimal NOT operation at the end. In view of the less precise alignment of the coordinate axes, this may seem a cheaper way of production of such states. However in order to achieve a good fidelity with optimal NOT operation, one needs multiple copies of anti-parallel classically correlated states at the beginning. We interpret this as yet another reason for an extra hidden classical correlation needed for production of parallel states $\rho^{\u\u}$. 
The higher correlation content of $\rho^{\u\u}$ shows itself in several aspects, we have shown this in detail for the set of one parameter Werner states, i.e. we show that the Werner state  
 $W^{\u\u}(t)$ is harder to produce, once produced, is more fragile against depolarizing noise, but is more efficient in an entanglement distribution protocol. 

\section*{Acknowledgment}
We would like to thank T. Abad for her contributions in the early stages of this project and her comments later on. We would like to thank Terry Rudolph and Lorenzo Maccone for their constructive comments and discussions. V. K. and L. M. thank Fabio Benatti and Roberto Floreanini for bringing reference \cite{LQU} to their attention and many subsequent discussions. V. K. and L. M  would like to thank The Abdus Salam ICTP for its hospitality and support under the Associate program in 2013.
\hspace{.3in}


\appendix

\section*{Appendix }\label{ApenA}

\setcounter{equation}{0}
\renewcommand{\theequation}{A\arabic{equation}}

In this appendix we detail the steps leading to equations (\ref{Fidelity}). The
aim is to calculate the fidelity of two $4\times 4$ matrices (\ref{E})
and (\ref{E'}).\\

Consider the state $\rho^{\u\u}$.
This state has a block structure in the form

 \be \label{AandB}
 \rho^{\u\u}=A\oplus B=\frac{1}{4}\left(\begin{array}{cc}1+t_3& t_1-t_2\\ t_1-t_2& 1+t_3\end{array}\right)\oplus \frac{1}{4}\left(\begin{array}{cc}1-t_3& t_1+t_2\\ t_1+t_2& 1-t_3\end{array}\right)
 \ee
 where $A$ and $B$ respectively denote the outer and the inner $2\times 2$ blocks.
We also write the local rotation operator $(I\otimes
R_z(\theta))$ in the same block structure form as: \be I\otimes
R_z(\theta)=\left(\begin{array}{cccc} e^{i\theta}&&&\\ &
e^{-i\theta} && \\ && e^{i\theta}& \\
&&&e^{-i\theta}\end{array}\right)=R_z(\theta)\oplus
R^\dagger_z(\theta). \ee This block structure will then give \be
{\rho^{\u\u}}^{\frac{1}{2}}=A^{\frac{1}{2}}\oplus
B^{\frac{1}{2}}\h
\rho^{\u\u}(\theta)=R_z(\theta)AR_z^{\dagger}(\theta)\oplus
R^{\dagger}_z(\theta)BR_z(\theta) \ee and greatly facilities
calculation of fidelity. We write \be
\sqrt{{\rho^{\u\u}}^{\frac{1}{2}}\rho^{\u\u}(\theta){\rho^{\u\u}}^{\frac{1}{2}}}=\sqrt{A^{\frac{1}{2}}R_z(\theta)AR^{\dagger}_z(\theta)A^{\frac{1}{2}}}\oplus
\sqrt{B^{\frac{1}{2}}R^{\dagger}_z(\theta)BR_z(\theta)B^{\frac{1}{2}}}
\ee where all matrices are now $2\times 2$ matrices. We now use
the following identity which is valid for any $2\times 2$ matrix
$M$:

\be tr\sqrt{M}=\sqrt{tr(M)+2\sqrt{det(M)}}. \ee This identity is
easily verified by diagonalizing the matrix. Using this identity,
we find \be F(W,
W(\theta))=\sqrt{tr(AR_z(\theta)AR^{\dagger}_z(\theta))+2
\det(A)}+\sqrt{tr(BR^{\dagger}_z(\theta)BR_z(\theta))+2 \det(B)}.
\ee Inserting $A$ and $B$ form (\ref{AandB}) into this formula we arrive at (\ref{Fidelity}).\\



\begin{thebibliography}{}

\bibitem{tele} C. H. Bennett, G. Brassard, C. Crepeau, R. Jozsa, A. Peres, and W. K. Wootters, Phys. Rev. Lett. \textbf{70}, 1895-1899 (1993).
\bibitem{dense} C. H. Bennett, S. J. Wiesner, Phys. Rev. Lett. \textbf{69}, 2881-2884 (1992).


\bibitem{entClassification1} W. Dur, G. Vidal, and J. I. Cirac, Phys. Rev. A \textbf{62}, 062314 (2000).
\bibitem{entClassification2} R. V. Buniy, and T. W. Kephart, J. Phys. A: Math. Theor. \textbf{45}, 185304 (2012); M. Aulbach, Int. J. Quantum Inform. \textbf{10}, 1230004 (2012). 
 

\bibitem{quantifyEnt} V. Vedral, M. B. Plenio, M. A. Rippin, and P. L. Knight, Phys. Rev. Lett. \textbf{78}, 2275 (1997).

\bibitem{Wootters} W. K. Wootters, Phys. Rev. Lett. \textbf{80}, 2245 (1998); For review see: R. Horodecki, P. Horodecki, M. Horodecki, K. Horodecki, Rev. Mod. Phys. \textbf{81}, 865–942 (2009).



\bibitem{entManipulation1} J. M. Raimond, M. Brune, and S. Haroche, Rev. Mod. Phys. \textbf{73}, 565 (2001).
\bibitem{entManipulation2} P. J. Shadbolt, M. R. Verde, A. Peruzzo, A. Politi, A. Laing, M. Lobino, J. C. F. Matthews, M. G. Thompson and J. L. O'Brien, Nature Photonics \textbf{6}, 45–49 (2012).


\bibitem{entDistribution1} J. I. Cirac, P. Zoller, H. J. Kimble, and H. Mabuchi, Phys. Rev. Lett. \textbf{78}, 3221 (1997).
\bibitem{entDistribution2} S. Perseguers, J. I. Cirac, A. Acin, M. Lewenstein, and J. Wehr, Phys. Rev. A \textbf{77}, 022308 (2008).

\bibitem{network2} S. Perseguers, M. Lewenstein,	A. Acin and J. I. Cirac, Nature Physics \textbf{6}, 539–543 (2010).



%
%
%
%




\bibitem{discord} H. Ollivier and W. H. Zurek, Phys. Rev. Lett. \textbf{88}, 017901 (2001); L. Henderson and V. Vedral, J. Phys. A \textbf{34}, 6899 (2001).

\bibitem{Groisman} B. Groisman, S. Popescu, and A. Winter, Phys. Rev. A \textbf{72}, 032317 (2005).


\bibitem{Modi} K. Modi, T. Paterek, W. Son, V. Vedral, and M. Williamson, Phys. Rev. Lett. \textbf{104}, 080501 (2010).

\bibitem{Gessner} M. Gessner, E.-M. Laine, H.-P. Breuer and J. Piilo, Phys. Rev. A \textbf{85}, 052122 (2012).

\bibitem{LQU} D. Girolami, T. Tufarelli, and G. Adesso, Phys. Rev. Lett. \textbf{110}, 240402 (2013).




\bibitem{Knill} E. Knill, and R. Laflamme, Phys. Rev. Lett. \textbf{81}, 5672 (1998). 

\bibitem{Datta} A. Datta, A. Shaji, and C. M. Caves, Phys. Rev. Lett. \textbf{100}, 050502 (2008).

\bibitem{Madhok} V. Madhok, and A. Datta, Phys. Rev. A \textbf{83}, 032323 (2011).

\bibitem{DiscInterpretation} D. Cavalcanti, L. Aolita, S. Boixo, K. Modi, M. Piani, and A. Winter, Phys. Rev. A \textbf{83}, 032324 (2011). 

\bibitem{Wang} L. Wang, J. H. Huang, J. P. Dowling, S. Y. Zhu, Quantum inf Process. \textbf{12}, 899-906 (2013).


\bibitem{RSP} C. H. Bennett, P. Hayden, D. W. Leung, P. W. Shor, A. Winter, IEEE Trans. Inform. Theory \textbf{12}, 56-74 (2005);
B. Dakic, Y. O. Lipp, X. Ma, M. Ringbauer, S. Kropatschek, S. Barz, T. Paterek, V. Vedral, A. Zeilinger, C. Brukner, and P. Walther, Nature Physics \textbf{8}, 666-670 (2012).

\bibitem{Werner} R. F. Werner, Phys. Rev. A \textbf{40}, 4277-4281 (1989).

\bibitem{WernerImportance} C. H. Bennett, G. Brassard, S. Popescu, B. Schumacher, J. A. Smolin, and W. K. Wootters, Phys. Rev. Lett. \textbf{76}, 722 (1996).

\bibitem{generation} B. P. Lanyon, P. Jurcevic, C. Hempel, M. Gessner, V. Vedral, R. Blatt, and C. F. Roos, Phys. Rev. Lett. \textbf{111}, 100504 (2013).

\bibitem{optNOT1} V. Buzek, M. Hillery, and R. F. Werner, Phys. Rev. A \textbf{60}, R2626(R) (1999).



\bibitem{RF-review} S. D. Bartlett, T. Rudolph, and R. W. Spekkens, Rev. Mod. Phys. \textbf{79}, 555 (2007).

\bibitem{Gisin} N. Gisin, and S. Popescu, Phys.  Rev. Lett. \textbf{83}, 432-435 (1999).


\bibitem{RF1} G. Chiribella, G. M. D'Ariano, P. Perinotti, and M. F. Sacchi, Phys. Rev. Lett \textbf{93}, 180503 (2004).

\bibitem{RF2} S. D. Bartlett, T. Rudolph, and R. W. Spekkens, and P. S. Turner, New J. Phys. \textbf{11}, 063013 (2009);
S. D. Bartlett, T. Rudolph, and R. W. Spekkens, Phys. Rev. Lett. \textbf{91}, 027901 (2003);
S. D. Bartlett, T. Rudolph, R. W. Spekkens, and P. S. Turner, New J. Phys. \textbf{8}, 58 (2006);
S. D. Bartlett, T. Rudolph, and R. W. Spekkens, Phys. Rev. A \textbf{70}, 032307 (2004).





\bibitem{qubitdiscord} Sh. Luo, Phys. Rev. A, \textbf{77} , 042303 (2008).


\bibitem{GeometricDiscord} B. Dakic, V. Vedral, and C. Brukner, Phys. Rev. Lett. \textbf{105}, 190502 (2010).

\bibitem{Bruss} A. Streltsov, H. Kampermann, and D. Bruss, Phys. Rev. Lett. \textbf{107}, 170502 (2011).

\bibitem{Abad} T. Abad, V. Karimipour, and L. Memarzadeh, Phys. Rev. A \textbf{86}, 062316 (2012).



\bibitem{Piani} M. Piani, Phys. Rev. A \textbf{86}, 034101 (2012).

\bibitem{skew} E. P. Wigner and M. M. Yanase, Proc. Natl. Acad. Sci. U.S.A. \textbf{49}, 910 (1963); S. Luo, Phys. Rev. Lett. \textbf{91}, 180403 (2003).


\bibitem{optNOT2} F. De Martini, V. Buzek, F. Sciarrino and C. Sias, Nature \textbf{419}, 815 (2002).
\bibitem{optNOT3} C. Simon, G. Weihs, and A. Zeilinger, J. Modern Optics \textbf{47}, 233 (2000).

\bibitem{discordRSP} G. L. Giorgi, Phys. Rev. A \textbf{88}, 022315 (2013).

\bibitem{Cubitt} T. S. Cubitt, F. Verstraete, W. Dur, and J. I. Cirac, Phys. Rev. Lett. \textbf{91}, 037902 (2003).

\bibitem{Peuntinger} C. Peuntinger, V. Chille, L. Mista, Jr. N. Korolkova, M. Fortsch, J. Korger, C. Marquardt, and G. Leuchs, Phys. Rev. Lett. \textbf{111}, 230506 (2013).



\end{thebibliography}
\end{document}